\documentclass[conference]{IEEEtran}
\IEEEoverridecommandlockouts

\usepackage{cite}
\usepackage{amsmath,amssymb,amsfonts}
\usepackage{algorithmic}
\usepackage{graphicx}
\usepackage{textcomp}
\usepackage{xcolor}
\def\BibTeX{{\rm B\kern-.05em{\sc i\kern-.025em b}\kern-.08em
    T\kern-.1667em\lower.7ex\hbox{E}\kern-.125emX}}

\usepackage{listings}
\newtheorem{problem}{Problem}
\usepackage{booktabs}

\usepackage{float} 
\usepackage[top=0.860in, bottom=1.2in, left=0.747in, right=0.65in]{geometry}

\usepackage{setspace}
\begin{document}

\title{Fast Online Digital Twinning on FPGA for Mission Critical Applications
}

\author{\IEEEauthorblockN{Bin Xu}
\IEEEauthorblockA{\textit{School of ECEE} \\
\textit{Arizona State University}\\
Tempe, USA \\
binxu4@asu.edu}
\and
\IEEEauthorblockN{Ayan Banerjee}
\IEEEauthorblockA{\textit{School of Computing and AI} \\
\textit{Arizona State University}\\
Tempe, USA \\
Ayan.Banerjee@asu.edu}
\and
\IEEEauthorblockN{Sandeep K.\,S. Gupta}
\IEEEauthorblockA{\textit{School of Computing and AI} \\
\textit{Arizona State University}\\
Tempe, USA \\
Sandeep.Gupta@asu.edu}
}

\maketitle

\begin{abstract}
Digital twinning enables real-time simulation and predictive modeling by maintaining a continuously updated virtual representation of a physical system. In mission-critical applications—such as mid-air collision avoidance—these models must operate online with extremely low latency to ensure safety. However, executing complex Model Recovery (MR) pipelines on edge devices is limited by computational and memory bandwidth constraints. This paper introduces a fast, FPGA-accelerated digital twinning framework that offloads key neural components—including gated recurrent units (GRU) and dense layers—to reconfigurable hardware for efficient parallel execution. Our system achieves real-time responsiveness, operating five times faster than typical human reaction time, and demonstrates the practical viability of deploying digital twins on edge platforms for time-sensitive, safety-critical environments.
\end{abstract}


\section{Introduction}

Digital twins are virtual models that mirror physical systems in real time using live data streams and simulations, enabling analysis, prediction, and control. According to Tao et al.~\cite{tao2018digital}, digital twins are increasingly deployed across domains such as manufacturing, aerospace, healthcare, and smart infrastructure, offering transformative capabilities in predictive maintenance, optimization, and decision support.

\begin{figure}[thbp]
\centering\includegraphics[width=0.85\columnwidth]{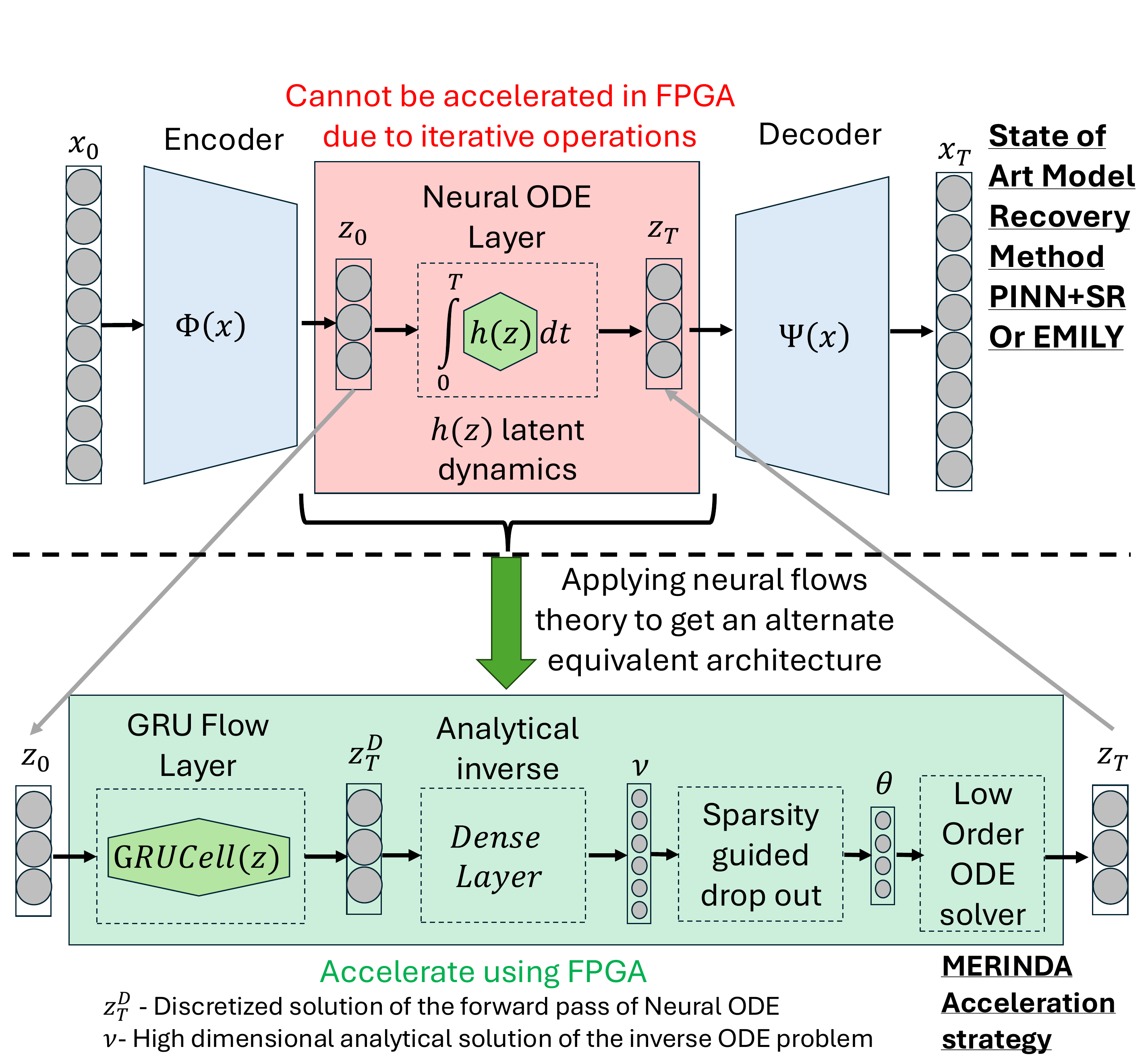}

    \caption{FPGA acceleration strategy using neural flow based equivalent architecture to neural ODEs.}
    \label{fig:FPGA} 
\end{figure}

At the core of digital twinning is data-driven predictive inference, which enables virtual models to simulate, monitor, and forecast system dynamics in real time. A key approach within this framework is physics-guided model recovery (MR)\cite{Banerjee2024}, where the underlying governing equations of system behavior are learned from real-world data—despite challenges such as sparse sampling, latent or unmonitored dynamics, and human-induced noise. Broadly, data-driven inference engines can be categorized into:
a) Generalized model learning (ML), where a high-dimensional neural network is trained directly to predict outputs from input data, and
b) Model recovery (MR), which aims to extract interpretable physical laws—e.g., through frameworks such as Physics-informed Neural Ordinary Differential Equations (PiNODE)\cite{PiNode}, sparse-regression PINNs (PINN+SR)\cite{chen2021physics}, or Extracting sparse Model from ImpLicit dYnamics (EMILY)\cite{pmlr-v255-banerjee24a,Banerjee2024} (Fig.~\ref{fig:FPGA}).

The strength of MR lies in its ability to detect subtle deviations from expected system behavior—such as actuator faults, sensor spoofing, or external disturbances—without relying on pre-defined fault categories. This makes MR particularly valuable in mission-critical autonomous systems (MCAS) environments, where rapid response and reliability are paramount. However, the deployment of MR algorithms in real-time settings remains computationally intensive, as many frameworks rely on solving ODEs, often through neural ODE-based architectures that are not optimized for low-latency inference.

To investigate this challenge in a practical safety-critical setting, we focus on the F8 Crusader aircraft model\cite{fan2020statistical}—a nonlinear dynamical system widely used in aerospace control and verification studies. Our objective is to apply MR techniques to detect collision-course anomalies and enable timely avoidance maneuvers. Prior studies have shown that human pilots typically require five seconds to respond to mid-air collision threats~\cite{cutler2017development}. Autonomous systems must react significantly faster, often in sub-second windows, to ensure safety in constrained airspace.

Achieving such rapid inference requires compute-efficient deployment of the digital twin's predictive components. In MCAS scenarios, systems face tight constraints on power, compute, and memory—particularly for onboard or edge devices. As a result, there is growing interest in accelerating inference models using reconfigurable hardware such as Field-Programmable Gate Arrays (FPGAs). These devices provide high-throughput, low-latency, and energy-efficient execution, making them ideal for edge-based AI systems~\cite{zhu2024edge, gunter2024apple}. Prior work has made significant progress in accelerating conventional ML models on FPGAs~\cite{ma2018optimizing, el2024fpga}. Existing FPGA-based MR frameworks\cite{xu2025accelerated,xu2025acceleratedECAI2025}, however, primarily target low-dimensional inputs in medical devices, with relatively little attention given to model scalability. This motivates our work on FPGA-based acceleration strategies specifically designed for MR-based digital twins operating on high-dimensional inputs.

To contextualize the challenge of accelerating model recovery (MR) architectures, it is helpful to contrast them with conventional time-series machine learning models. The widespread adoption of FPGA accelerators for time-series ML predictors can be attributed to their highly regular and well-understood architecture. These models typically consist of an input embedding layer followed by a stack of recurrent neural network (RNN) layers that compute discretized temporal dynamics. Their forward pass reduces to structured matrix multiplication operations, while nonlinear functions such as sine or cosine can be efficiently computed using specialized FPGA components like CORDIC~\cite{Cao20Cordic}. The backward pass follows standard gradient-based pipelines, further simplifying their suitability for hardware acceleration.

In contrast, neural ODE-based architectures such as ODENet~\cite{Wtanabe} and NODE~\cite{CaiFPGA} present greater challenges for acceleration. These works have primarily focused on fixed-depth, static architectures, which limit their flexibility and applicability to general MR frameworks. While they offer valuable insights into hardware-aware ODE acceleration, they fall short of supporting dynamic architectures like PiNODE, PINN+SR, or EMILY, where the number of NODE layers may vary depending on the application or system complexity.

As shown in Fig.~\ref{fig:FPGA}, a significant portion of EMILY (a baseline comparator in our study) is composed of NODE cells~\cite{chen2018neural}. The forward pass of a NODE cell involves solving a high-dimensional ODE using numerical solvers with tight precision requirements. One of the fundamental challenges in accelerating MR frameworks lies in the inherently iterative and adaptive nature of this computation. Recent studies have attempted to accelerate ODE solvers for standalone, fixed-coefficient equations on FPGA~\cite{stamoulias2017high,ebrahimi2017evaluation}, but these designs are not suitable for PiNODE-like settings, which require solving a large set of ODEs with varying, input-dependent model coefficients.

In this paper, we utilize the theory of neural flows~\cite{bilovs2021neural} to obtain an alternative neural structure called MERINDA, model recovery in dynamic architectures, that is equivalent to the NODE layers used in EMILY, PiNODE or PINN+SR and more amenable for acceleration in an FPGA. MERINDA replaces the NODE layer with a layer of invertible functions designed using a combination of gated recurrent units (GRU) and a dense layer of neurons with nonlinear activation functions. 
GRU~\cite{shiri2023comprehensive} are a type of recurrent neural network (RNN) architecture that introduces gating mechanisms to control the flow of information over time. Compared to traditional RNNs or LSTMs, GRUs are computationally efficient and require fewer parameters~\cite{chung2014empirical}, making them favorable for deployment on resource-constrained platforms such as FPGAs.

The primary contributions of this paper are as follows:

\noindent\textbf{1.} An empirical evaluation of MERINDA against state-of-the-art model recovery (MR) approaches—specifically EMILY and PINN+SR—across four benchmark nonlinear dynamical systems, including two real-world and two simulated datasets, demonstrating its MR accuracy.

\noindent\textbf{2.} A detailed analysis of FPGA resource utilization and throughput across varying dimensional configurations of the F8 Crusader system, along with performance metrics after hardware-specific optimization.

\noindent\textbf{3.} A comparative study between MERINDA and GPU-based implementations (e.g., CUDA) to quantify the acceleration benefits.

\section{Theoretical background}
This section presents the basics of MR and establishes approximate equivalence of neural flow architecture with NODE.
\subsection{Basics of Model Recovery}
The main goal of MR is akin to an auto-encoder (Fig. \ref{fig:FPGA}), where given a multivariate time series signal $X(t)$, the aim is to find a latent space representation that can be used to reconstruct an estimation $\Tilde{X}(t)$ with low error. It has the traditional encoder $\phi(t)$ and decoder ($\Psi(t)$) of an autoencoder architecture. MR represents the measurements $X$ of dimension $n$ and $N$ samples, as a set of nonlinear ordinary differential equation model in (\ref{eqn:Model}). 
\begin{equation}
    \label{eqn:Model}
    \scriptsize
    \dot{X} = h(X,U,\theta),
\end{equation}
where $h$ is a parameterized nonlinear function, $U$ is the $m$ dimensional external input, and $\theta$ is the $p$ dimensional coefficient set of the nonlinear ODE model.

\noindent{\bf Sparsity:} An $n$-dimensional model with $M^{th}$ order nonlinearity can utilize $\binom{M+n}{n}$ nonlinear terms. A sparse model only includes a few nonlinear terms $p << \binom{M+n}{n}$. Sparsity structure of a model is the set of nonlinear terms used by it.

\noindent{\bf Identifiable model:} A model in (\ref{eqn:Model}) is identifiable~\cite{verdiere2019systematic}, if $\exists$ time $t_I > 0$, such that $\forall \theta, \Tilde{\theta} \in \mathcal{R}^p$:
\begin{equation}
    \label{eqn:Ident}
    \scriptsize
    \forall t \in [0,t_I], f(X(t),U(t),\theta) = f(X(t),U(t),\Tilde{\theta}) \implies \theta = \Tilde{\theta}. 
\end{equation}

\noindent Eqn. \ref{eqn:Ident} effectively means that a model is identifiable if two different model coefficients do not result in identical measurement $X$. In simpler terms this means $\forall \theta_i \in \theta, \frac{dX}{d\theta_i} \neq 0$. In this paper, we assume that the underlying model is identifiable.

\begin{problem}[Sparse Model Recovery]\label{prob:Problem} Given $N$ samples of measurements $X$ and inputs $U$, obtained from a sparse model in Eqn. \ref{eqn:Model} such that $\theta$ is identifiable, recover $\Tilde{\theta}$ such that for $\Tilde{X}$ generated from $f(X,U,\Tilde{\theta})$, we have $||X - \Tilde{X}|| \leq \epsilon$, where $\epsilon$ is the maximum tolerable error.
\end{problem}

\noindent{\bf Role of NODE:} Both EMILY~\cite{pmlr-v255-banerjee24a} and PINN+SR~\cite{chen2021physics} utilize a layer of NODE cells in order to integrate the underlying nonlinear ODE dynamics. NODE cell's forward pass is by design the integration of the function $h$ over time horizon $T$ with $N$ samples (Fig. \ref{fig:FPGA}). This effectively requires an ODE solver in each cell of the NODE layer:
\begin{equation}
\scriptsize
z(t) = \int\limits^T_0{h(z,u,\theta)dt}, 
\end{equation}
where $z \in Z$ and $u \in U$ are each cells output and input. 

The results are then used further in the EMILY or PINN+SR pipeline to extract the accurate underlying nonlinear ODE model.

\begin{figure*}[t]
    \centering
    \includegraphics[width=0.75\textwidth]{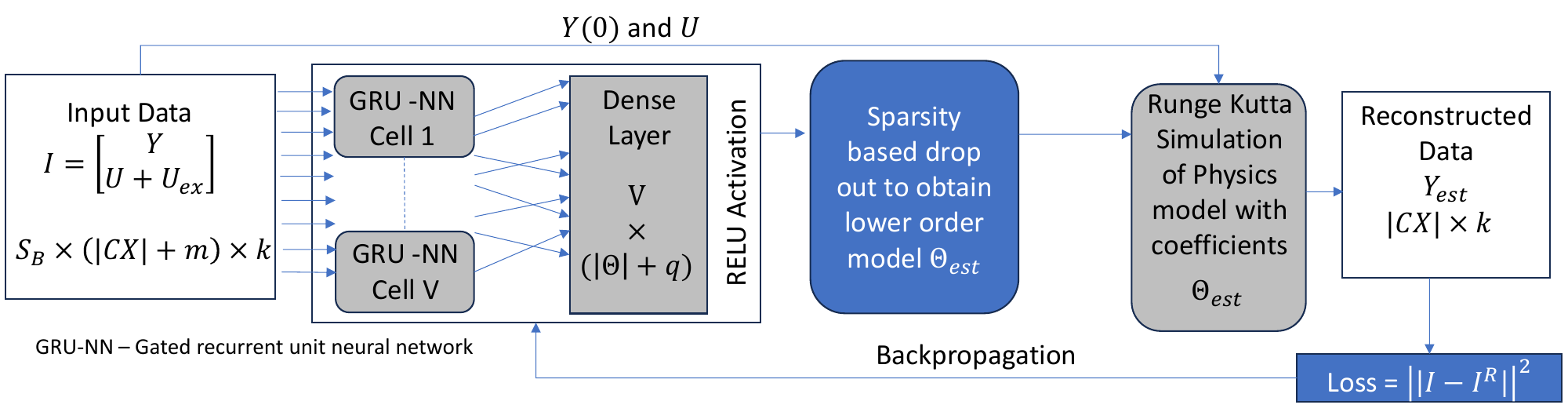}
    \caption{MERINDA: Gated recurrent unit (GRU) NN-based MR architecture.}
    \label{fig:Approach}
\end{figure*}

\subsection{Neural flows and equivalent architectures to NODE}

According to the theory of neural flows~\cite{bilovs2021neural}, the node layer can be replaced by an approximate solution to $F(t) \approx Z(t)$ in discretized form using recurrent nerual network architectures such as GRU provided that following conditions are satisfied:

\begin{scriptsize}
\begin{equation}
    F(0,u) = Z(0,u), \text{(initial condition), and  }
    F(t,u) \text{  is invertible}.
\end{equation}
\end{scriptsize}

Bilovs et al.~\cite{bilovs2021neural} show that $F(t,u)$ can be achieved by replacing the original NODE layer by a GRU layer. However, the GRU layer does not statisfy by the invertible condition. The authors in~\cite{bilovs2021neural} suggest the usage of a dense layer since it acts as a universal approximator of nonlinear functions and hence can also act as the inversion of the function $F(t,u)$.  

MERINDA further enhances the equivalent architecture proposed in~\cite{bilovs2021neural} by further pruning the dense layer as shown in Fig. \ref{fig:Approach}. The main idea is to further reduce the dense layer structure by utilizing the inherent sparsity in the data.
Given the definitions of identifiability and sparsity, we now discuss our full architecture that is equivalent to PiNODE



\section{MERINDA Architecture}

\subsection{GRU NN-based MR architecture}
In our approach (Fig. \ref{fig:Approach}), we extend gated recurrent unit neural network (GRU-NN) to obtain advanced neural structure MERINDA that can solve the model recovery problem. The forward pass of GRU-NN structure expresses the coefficients of the model as a nonlinear function of the outputs $Y$ and inputs $U$ of the model. The measurements of $Y$, can be used to convert the set of implicit dynamics to an overdetermined system of equations that are nonlinear in terms of the model coefficients. As such an over-determined system of equation may have no solution unless either some equations are rejected or are expressed as linear superposition of other equations. To search for a set of consistent equations to estimate model coefficient, a dense layer is utilized. The search process of the dense layer is guided by a loss function (\textit{ODE loss}) that computes the mean square error between the estimated $Y_{est}$ using an ODE solver $\mathbf{SOLVE}(Y(0),\Theta, U)$ and the ground truth measurements $Y$.


The advanced neural architectures for model recovery in Fig. \ref{fig:Approach} is implemented by extending the base code available in~\cite{liquid-time-constant-networks}. We extract the training data consisting of temporal traces of $Y$, and $U$ . $Y$ is sampled at least at the Nyquist rate for the application, and $U$ has the same sampling rate as $Y$. The resulting training data is then divided into batches of size $S_B$. This forms a 3D tensor of size $S_B \times |Y|+m \times k$. 

Each batch is passed through the {\it GRU-NN} network with $V$ nodes, resulting in $V$ hidden states. A dense layer is then employed to transform these $V$ hidden states into $p = |\Theta|$ model coefficient estimates and $q$ input shift values. The dense layer is a multi-layer perceptron with ReLU activation function for the model coefficient estimate nodes, whose outputs are the estimated model coefficients. The dense layer converts the $V$ dimensional hidden layer outputs to $M+|X| \choose |X|$ which is the number of nonlinear terms that can be used for an $M$th order polynomial. A dropout rate of $|\Theta|$ is used so that the final number of output layers with non-zero activation is $|\Theta|$. The model coefficient estimates and the initial value $Y(0)$ is passed through an ODE solver to solve the nonlinear dynamical equations with the coefficients $\Theta_{est}$, initial conditions $Y(0)$ and inputs $U$. The Runge Kutta integration method is used in the ODE solver, which gives $Y_{est}$. In the backpropagation phase the network loss is appended with ODE loss, which is the mean square error between the original trace $Y$ and the estimated trace $Y_{est}$.

\subsection{FPGA Architecture and Optimization}
An FPGA (Field-Programmable Gate Array) is a reconfigurable semiconductor device that enables developers to implement custom digital circuits directly in hardware~\cite{maxfield2004design}. Unlike fixed-function processors, FPGAs consist of an array of Configurable Logic Blocks (CLBs), Look-Up Tables (LUTs) for implementing combinational logic, flip-flops and registers for sequential logic, and programmable interconnects~\cite{kuon2007measuring}. FPGAs also incorporate on-chip memory resources, such as Block RAM (BRAM) and UltraRAM (URAM), as well as Digital Signal Processing (DSP) slices optimized for arithmetic-intensive operations. 

One of the primary challenges in FPGA design lies in efficiently mapping high-level algorithms onto limited hardware resources while maximizing performance. Loop-carried dependencies—such as Read-After-Write (RAW), Write-After-Read (WAR)—can inhibit effective pipelining, and reduce throughput. In addition to control hazards, memory access patterns pose a significant design challenge. FPGAs feature a hierarchical memory system including block RAM (BRAM), Look-Up Tables (LUTs), and Flip-Flops (FF), all of which must be judiciously partitioned and scheduled to avoid access bottlenecks and ensure data locality.

\begin{figure}[h!]
\centering
\includegraphics[width=\columnwidth]{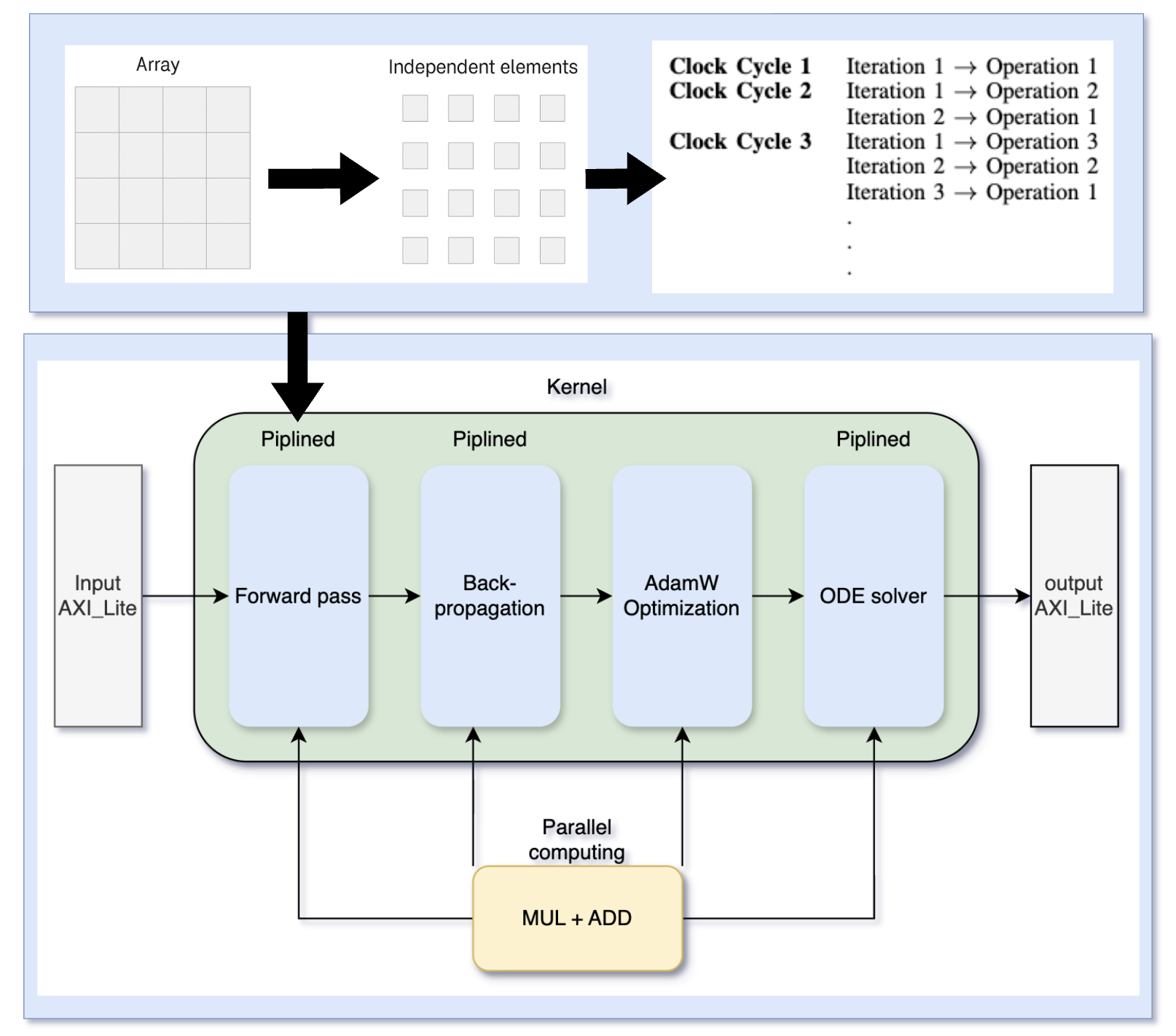} 
\caption{After the input data is partitioned, then pipeline can be applied to compute.} 
\label{fig:partition}
\end{figure}

In our design, illustrated in Fig.~\ref{fig:partition}, we address these challenges through two key techniques: \textit{array partitioning} and \textit{loop pipelining}, both guided by high-level synthesis (HLS) directives. The FPGA kernel interfaces with the processor using an AXI4-Lite protocol, after which input data is transferred to on-chip memory. We apply full array partitioning using the directive \texttt{\#pragma HLS ARRAY\_PARTITION complete}, which instructs the HLS compiler to map each element of the input array to an independent storage resource—such as a dedicated register or BRAM segment. This partitioning strategy eliminates inter-element memory conflicts and enables parallel access to the data elements.

As shown in Fig.\ref{fig:partition}, individual elements like \texttt{array[0][0]} through \texttt{array[0][3]} are independently accessible and routed to distinct computational units, allowing multiple operations to proceed concurrently. This fine-grained parallelism directly mitigates the memory bottlenecks described earlier, as each operation now operates on separate physical storage units. Additionally, it maximizes utilization of on-chip memory resources by spreading the access load across registers, BRAM, and LUTs.

We then construct a fully parallelized model recovery pipeline on the FPGA. All major computational stages—including the forward pass, backpropagation, and loss computation—are pipelined using \texttt{\#pragma HLS PIPELINE II=1}. Once the inputs are partitioned and loop-carried dependencies are removed, this setup achieves an initiation interval (II) of 1, allowing a new iteration to begin every clock cycle. This significantly boosts throughput and latency performance. There can be the violation of loop dependency in the simulation. In order to eliminate RAW and WAR hazards, we need to test \texttt{\#pragma HLS PIPELINE II=2} or \texttt{\#pragma HLS PIPELINE II=3}, which means a new iteration begins in every 2 cycles and 3 cycles. If there is no time violation in the simulation, it means there are no RAW and WAR hazards. 
However, more cycles mean more latency in the pipeline of computation. 

In the presence of data dependencies, such as in the sequentially linked operations of the GRU forward pass (Operations 1 to 3) and backpropagation, pipelining allows for overlapping execution. Specifically, operations 1--3—outlined in the forward pass code—exhibit loop-carried dependencies, such as between \texttt{z[i]} and \texttt{rz\_concat[i]}. With proper pipelining and no write-read hazards, the next operation can begin in the following cycle, thereby enabling deep pipelining across dependent computations.

\subsection*{Operation 1: Compute Reset and Update Gates}
\begin{lstlisting}[language=C++, basicstyle=\ttfamily\small, frame=single]
for i in 0 to H-1:
    z_sum = bias_z[i]
    for j in 0 to H + V - 1:
        z_sum += Wz[i][j] * concat[j]
    z[i] = sigmoid(z_sum)
\end{lstlisting}

\subsection*{Operation 2: Apply reset gate to previous hidden state}
\begin{lstlisting}[language=C++, basicstyle=\ttfamily\small, frame=single]
for i in 0 to H-1:
    rz_concat[i] = r[i] * a_prev[i]
\end{lstlisting}

\subsection*{Operation 3: Compute Candidate Activation}
\begin{lstlisting}[language=C++, basicstyle=\ttfamily\small, frame=single]
for i in 0 to H-1:
    cc_sum = bias_c[i]
    for j in 0 to H + V - 1:
        cc_sum += Wa[i][j] * rz_concat[j]
    c_t[i] = tanh(cc_sum)
\end{lstlisting}

\section{Evaluation and Results}




\subsection{Implementation Details}\label{AA}
To evaluate the performance of the FPGA, we perform experiments on the mobile GPU and FPGA with the same hidden layer size, epoch size and model dimension size. The performance of the mobile GPU is set as a baseline. 

\paragraph{Mobile GPU Platform} Our experiments are conducted on the NVIDIA Jetson Orin Nano Developer Kit, which features a 6-core Arm Cortex-A78AE CPU and 8~GB of LPDDR5 memory. The integrated GPU is based on the NVIDIA Ampere architecture, equipped with 1024 CUDA cores and 32 Tensor Cores.

We implemented our GPU simulation using CUDA C++, which offers near-zero runtime overhead compared to Python's interpreted execution model. Unlike Python, CUDA C++ enables fine-grained control over thread management, shared memory reuse, and global memory access patterns, allowing for highly optimized low-level execution~\cite{oden2020lessons}. In our evaluations, the CUDA C++ implementation achieved over 10× speedup relative to its Python version.

\paragraph{FPGA Platform}
For the FPGA platform, the experiments were performed on Zynq UltraScale+ MPSoC with Quad ARM Cortex-A53, which includes 252K LUTs and 504K Flip-Flops. The GRU cell was built from scratch in C++ using High-Level Synthesis (HLS) on AMD’s Vitis tool. The whole model recovery is simulated on Vitis. The execution time is based on \textbf{latency (cycles)} multiplied by the \textbf{clock period (ns)}.

\subsection{Results}\label{AA}

\begin{table}[h!]
\centering
\scriptsize
\caption{Comparison between MERINDA and SOTA MR techniques EMILY and PINN+SR using reconstruction MSE. Errors are absolute values; numbers in parentheses indicate standard deviation.}
\begin{tabular}{|l|c|c|c|}
\hline
\textbf{System} & \textbf{EMILY} & \textbf{PINN+SR} & \textbf{MERINDA} \\
\hline
Lotka Volterra      & 0.03 (0.02)     & 0.05 (0.03)     & 0.03 (0.018) \\
Chaotic Lorenz      & 1.7 (0.6)       & 2.11 (1.4)      & 1.68 (0.4) \\
F8 Crusader          & 4.2 (2.1)       & 6.9 (4.4)       & 5.1 (2.2) \\
Pathogenic Attack   & 14.3 (12.1)     & 21.4 (5.4)      & 15.1 (10.2) \\
\hline
\end{tabular}
\vspace{1.3pt}
\label{tbl:SMRFM}
\end{table}

\paragraph{Accuracy Comparison of MERINDA with EMILY and PINN+SR}
We evaluate the accuracy of MERINDA using the mean squared error (MSE) metric across standard benchmark examples from~\cite{kaiser2018sparse}. Table~\ref{tbl:SMRFM} reports the reconstruction errors for MERINDA, along with published results for EMILY~\cite{pmlr-v255-banerjee24a} and PINN+SR~\cite{robinson2022physics}. The comparison shows that MERINDA achieves comparable or improved accuracy relative to these state-of-the-art model recovery techniques while offering additional benefits in efficiency and hardware acceleration.

\begin{figure}[thbp]
\centering
\includegraphics[width=0.9\columnwidth]{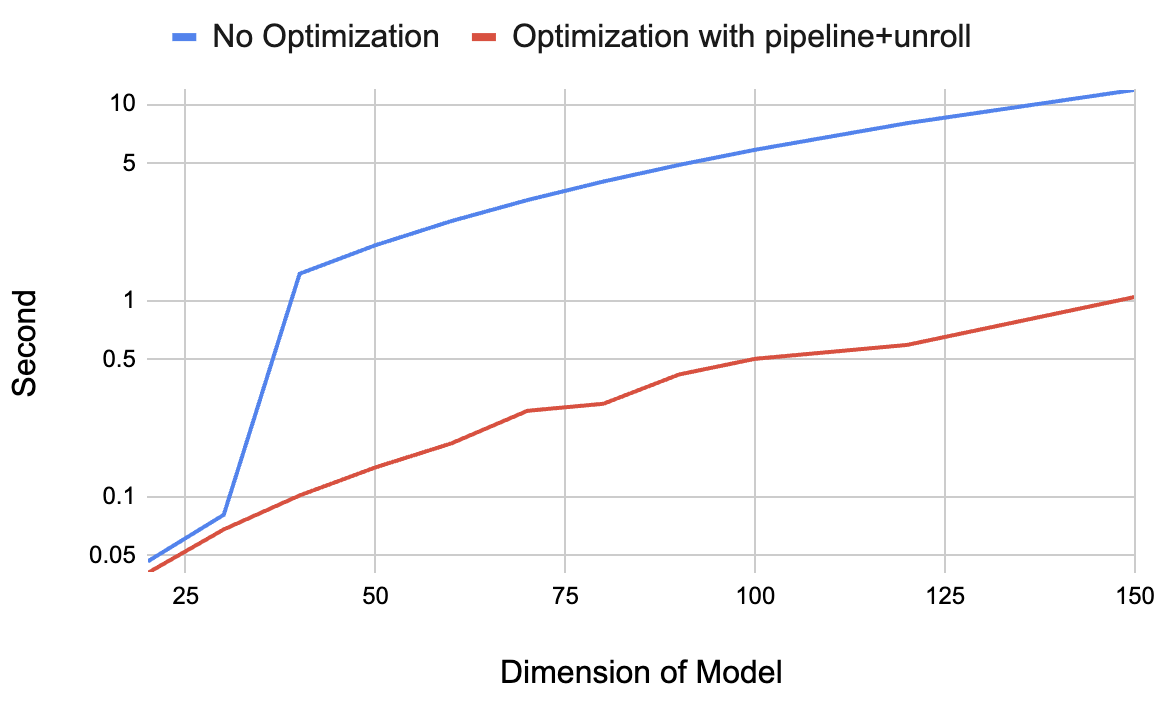} 
\caption{Impact of Optimization on Model Recovery Execution Time.}
\label{fig:performance_optimization_vs_no_optimization}
\end{figure}

\paragraph{Performance of MERINDA on F8 Crusader} 
The Fig. \ref{fig:performance_optimization_vs_no_optimization} compares the model recovery time on FPGA for varying model dimensions, with and without hardware-level optimization. The x-axis denotes the dimensionality of the F8 Crusader model (Eqs. 7, 8 and 9 from\cite{fan2020statistical}), and the y-axis shows the total execution time in seconds (log-scaled).

The blue curve represents the baseline implementation with no optimization, which exhibits a steep and nonlinear increase in runtime as the model dimension grows. This is primarily due to increased loop latency and limited parallelism in the unoptimized pipeline.
In contrast, the red curve shows the performance of the optimized implementation, which employs both loop pipelining and unrolling. These optimizations reduce latency by allowing concurrent computation and minimizing memory access bottlenecks.
Notably, at model dimension 150, the optimized design achieves more than a 10× speedup compared to the unoptimized version (1.04s vs. 11.84s), highlighting the effectiveness of hardware optimization in scaling up real-time model recovery.

\begin{table}[h!]
\centering
\scriptsize
\caption{Comparison of resource usage and execution time between FPGA and GPU for increasing F8 Crusader model dimension. BRAM(KB), FPGA(s), gpu(s)}
\begin{tabular}{|c|c|c|c|c|c|c|}
\hline
\textbf{Size} & \textbf{Cycles} & \textbf{LUT} & \textbf{DSP} & \textbf{BRAM} & \textbf{FPGA} & \textbf{GPU} \\
\hline
20  & 17,019   & 314,433   & 2,419  & 95  & 0.0408  & 0.323  \\
30  & 28,336   & 463,953   & 3,745  & 135 & 0.0680  & 0.387  \\
40  & 42,255   & 616,316   & 5,003  & 247 & 0.1014  & 0.400  \\
50  & 58,754   & 745,437   & 6,119  & 297 & 0.1410  & 0.428  \\
60  & 77,875   & 891,062   & 7,466  & 347 & 0.1869  & 0.494  \\
70  & 114,082  & 806,340   & 7,276  & 397 & 0.2738  & 0.535  \\
80  & 123,915  & 1,194,348 & 9,826  & 447 & 0.2974  & 0.619  \\
90  & 174,862  & 1,038,178 & 9,271  & 497 & 0.4197  & 0.990  \\
100 & 210,052  & 1,128,589 & 10,139 & 547 & 0.5041  & 1.051  \\
120 & 247,195  & 1,779,201 & 14,831 & 647 & 0.5933  & 1.244  \\
150 & 434,003  & 1,690,814 & 15,270 & 797 & 1.0416  & 1.541  \\
\hline
\end{tabular}
\label{tab:f8_fpga_gpu_grid}
\end{table}

Table~\ref{tab:f8_fpga_gpu_grid} presents the resource utilization and execution time of the model recovery process on FPGA and GPU as the F8 Crusader model dimension increases. As expected, both execution time and hardware resource consumption (LUTs, DSPs, BRAM) grow with model complexity. Compared to GPU implementation, the FPGA implementation shows lower latency because more weights are loaded onto registers such as LUT for computation. With the pipeline optimization, the computation on FPGA can reuse data on shared memory.  


Table~\ref{tab:fpga_optimization_results} compares three FPGA configurations using dimension 30 as a reference point: a baseline design with no optimization, a partially optimized version with only unrolling in the loop, and a fully optimized version combining both loop pipelining and unrolling. The fully optimized design achieves the lowest latency (0.0680 seconds), a 1.4$\times$ speedup over the unoptimized version. It also exhibits more efficient resource utilization across DSPs and BRAM, validating the effectiveness of `\texttt{\#pragma HLS PIPELINE}` and `\texttt{ARRAY\_PARTITION}` in achieving high throughput on FPGAs. From the resource usage shown in the Table\ref{tab:fpga_optimization_results}, fully optimization store more data and compute in the LUT and FF.


\begin{table}[h!]
\centering
\scriptsize
\caption{FPGA resource utilization and execution time for different optimization strategies (dimension = 30).}
\begin{tabular}{|l|r|r|r|r|r|}
\hline
\textbf{Configuration} & \textbf{Cycles} & \textbf{LUT} & \textbf{DSP} & \textbf{BRAM(KB)} & \textbf{Time(s)} \\
\hline
No Optimization         & 40,390  & 219,238  & 372   & 136 & 0.0966 \\
Unroll                  & 34,991  & 366,741  & 1,847 & 154 & 0.0840 \\
Pipeline + Unroll       & 28,336  & 463,953  & 3,745 & 135 & 0.0680 \\
\hline
\end{tabular}
\vspace{1.5pt}
\label{tab:fpga_optimization_results}
\end{table}

\section{Discussion}
In Table~\ref{tab:f8_fpga_gpu_grid}, we compare the timing of our FPGA implementation with a 13W mobile GPU, selected for its relevance in edge AI. For low-dimensional F8 Crusader models, the FPGA achieves superior speed due to its efficient parallelism and low-power design. However, beyond dimension 150, performance degrades as limited on-chip memory forces frequent off-chip access, introducing latency. In contrast, the mobile GPU handles high-dimensional models more effectively, thanks to greater on-chip memory and higher memory bandwidth. This reveals a key trade-off: FPGAs are optimal for compact, parallel workloads, while GPUs excel in memory-intensive scenarios.

During synthesis, we observed that Vitis HLS can report hardware resource utilization exceeding the physical limits of the target FPGA board. For example, at higher model dimensions (e.g., dimension 150), the estimated usage of LUTs and BRAMs surpassed the available resources on the device. It appears that Vitis sometimes overestimates utilization due to conservative analysis during high-level synthesis, particularly when aggressive unrolling or partitioning is used. In practice, actual post-implementation resource usage is often lower after placement, routing, and optimization steps are completed.

Since we could not fully identify the root cause of this discrepancy within the scope of this study, we include this disclaimer:
Resource utilization values reported by Vitis HLS for high-dimensional configurations may exceed the target device limits, so resource estimates should be interpreted cautiously, and actual deployability should be confirmed through on-board testing. This highlights a broader need for more reliable resource estimation in HLS workflows and calls for careful validation when designing edge-deployable FPGA systems for mission-critical applications.

\section{Conclusions}
This work addresses the urgent need for fast, reliable, and interpretable model recovery (MR) in mission-critical autonomous systems (MCAS), where real-time anomaly detection and response are essential for safety. We proposed MERINDA, an FPGA-accelerated model recovery framework capable of recovering governing dynamics from data using architectures equivalent to PiNODE and EMILY.

By targeting the F8 Crusader benchmark and simulating the model on FPGA, we demonstrated MERINDA's capacity to support sub-second reaction times—far outperforming the 5-second human pilot baseline cited in prior literature~\cite{cutler2017development}. Unlike traditional CPU or GPU implementations, MERINDA leverages HLS-based pipeline and parallelism strategies to significantly reduce execution latency, while maintaining accuracy and interpretability.

\section*{Acknowledgement}
This work was partially supported by DARPA (AMP, N6600120C4020; FIRE, P000050426), the NSF (FDT-Biotech, 2436801), and the Helmsley Charitable Trust (2-SRA-2017-503-M-B).

\end{document}